# On the stability of a very dense deuterium-tritium plasma ball

Yuri Kornyushin

*Maître Jean Brunschvig Research Unit, Chalet Shalva, Randogne, 3975-CH*

Email: jacqie@bluewin.ch

General conditions of a stability of a very dense deuterium-tritium plasma ball are discussed. It is shown that the decrease in the size of a plasma ball (increase in the plasma density) can be achieved somewhat easier when the temperature and the pressure in the plasma ball are kept high enough for all the particles, the nuclei and the delocalized electrons, to be described by classical statistics.

Let us consider a very dense neutral deuterium-tritium plasma ball of a radius $R$, consisting of $N$ nuclei and $N$ delocalized electrons. The Gibbs free energy of the plasma ball is

$$\Phi = PV + K + U - TS, \qquad (1)$$

where $P$ is the external pressure, $V = (4\pi/3)R^3$ is the volume of a ball, $K$ is the kinetic energy of the particles, $U$ is the potential electrostatic energy, $T$ is the absolute temperature, and $S$ is the entropy. When the delocalized electrons are degenerate, the kinetic energy of them in the ball of a radius $R$ is $K_e = 1.105 N^{5/3} \hbar^2 / m R^2$ [1]. The potential electrostatic energy of a neutral plasma ball consists of two terms. One of them, $-0.75 e^2 N g$, describes a negative contribution of the screening to the potential electrostatic energy [1] (here $e$ is elementary charge, and $1/g$ is the screening radius). The origin of this contribution is a relaxation of the delocalized electrons to lower energy state of screening of the ions. When delocalized electrons relax to more equilibrium state, screening the ions, the electrostatic energy of a sample decreases. Screening cuts off the long-range electrostatic field of the ions, that is, leads to the decrease in the electrostatic energy of a sample. This problem was considered in [2]. In [2] the ionization energy of impurities in semiconductors was discussed. The effective ionization energy $W_e$ was calculated in [2] for the case when the radius of a ground state is much smaller than the screening radius. The result was as follows: $W_e = W_0 - e^2 N g$ (here $W_0$ is the initial ionization energy). This corresponds to taking into account only the first term in Eq. (29) in [1].

The other term of the electrostatic energy, $-1.2 e^2 N / R$, describes (in the model of homogeneously distributed in a ball charge of every particle) a negative contribution, coming from the absence of the self-action in quantum mechanics. Potential energy $U(\mathbf{r})$ in Schrödinger equation for elementary charged particle does not include the potential energy of the interaction of the considered elementary charged particle with the field produced by the same particle. This means that in quantum mechanics the action of the elementary charged particle on itself, that is a self-action, is not taken into account. Averaging Schrödinger equation quantum-mechanically, one can obtain that the energy of a particle consists of its kinetic energy and its potential energy in the external (relative to the regarded particle) potential field. So one can see that the energy of the regarded particle do not include the potential energy of the field created by the particle. This means that any elementary charge $e$ in any state has no electrostatic potential energy of its own in quantum mechanics, see e.g. [3]. From this follows that the electrostatic potential energy of the regarded neutral plasma ball is $-1.2e^2N/R$.

For degenerate electrons $g$ is proportional to $1/R^{1/2}$ [1,4]. For the classical ones $g$ is proportional to $1/R^{3/2}$ [4]. For $N$ classical particles, as is well known, the negative contribution of the entropy to Gibbs free energy is $-3kTN\ln(R/R_0)$, where $R_0$ is some parameter.

So, when electrons are degenerate, the negative part of the Gibbs free energy (one term is proportional to $1/R$, the other one is proportional to $1/R^{1/2}$, the third one is proportional to $\ln(R/R_0)$) cannot compensate for the positive term of the kinetic energy, proportional to $1/R^2$, in the case of a decrease in the ball radius $R$. So when temperature $T$ and the plasma density $N/V$ correspond to the delocalized electrons being degenerate, the further decrease in the plasma ball size is somewhat difficult.

On the contrary, when the temperature is high enough for all $2N$ particles (the nuclei and the delocalized electrons) to be classical, the kinetic energy of all the particles is $3kTN$. It does not depend on the ball radius $R$. In this case further decrease in the plasma ball size can be somewhat easier. Decrease in the plasma ball size, however, leads to the increase in the plasma density and brings the plasma ball out of the classical statistics, when the plasma density increases. This makes the further decrease in the plasma ball size more difficult. To keep the process of the plasma ball size decreasing easier, the temperature and the pressure should be kept high enough, when classical statistics is relevant. The process of decreasing the plasma ball size can result in the beginning of nuclear fusion, when the plasma density becomes high enough. However this requires extremely high temperature and pressure.

It should be noted that the process of decreasing of the regarded plasma ball size, when the particles are essentially degenerate, is possible [5,6]. Here it was discussed how to keep this process of the plasma densification easier.

## О стабильности очень плотной дейтериево-тритиевой плазмы

Ю. В. Корнюшин

*Maître Jean Brunschvig Research Unit, Chalet Shalva, Randogne, 3975-CH*
Email: jacqie@bluewin.ch

Обсуждаются общие условия стабильности очень плотной дейтериево-тритиевой плазмы. Показано, что уменьшение размера плазменного шара (увеличение плотности плазмы) можно достигнуть легче при обеспечении достаточно высоких температур и давлений, при которых все частицы (как ядра, так и делокализованные электроны) описываются классической статистикой.

Рассмотрим очень плотный шар нейтральной дейтериево-тритиевой плазмы радиуса $R$, содержащий $N$ ядер и $N$ делокализованных электронов. Термодинамический потенциал рассматриваемого шара

$$\Phi = PV + K + U - TS. \qquad (1)$$

Здесь $P$ обозначает внешнее давление, объём шара $V = (4\pi/3)R^3$, $K$ – кинетическая энергия частиц, $U$ – потенциальная электростатическая энергия, $T$ – абсолютная температура и $S$ – энтропия. Если делокализованные электроны вырождены, их кинетическая энергия в шаре радиуса $R$ есть $K_e = 1.105 N^{5/3} \hbar^2 / mR^2$ [1]. Потенциальная электростатическая энергия нейтральной плазмы состоит из двух слагаемых. Одно из них, $-0.75 e^2 Ng$, представляет отрицательный вклад экранирования в потенциальную электростатическую энергию [1] (здесь $e$ – элементарный заряд, а $1/g$ – радиус экранирования). Это слагаемое происходит от релаксации делокализованных электронов к более равновесному экранирующему ионы состоянию. При этом электростатическая энергия образца уменьшается. Экранирование обрезает дальнодействующее электростатическое поле ионов и обуславливает уменьшение электростатической энергии образца. Эта задача была рассмотрена в [2]. В [2] обсуждалась энергия ионизации примесей в полупроводнике. В [2] была вычислена эффективная энергия ионизации примеси $W_e$ для случая когда радиус основного состояния значительно меньше радиуса экранирования. Было получено что $W_e = W_0 - e^2 ng$ (здесь $W_0$ – энергия ионизации без учёта экранирования). Этот результат соответствует учёту только первого слагаемого в уравнении (29) в [1].

Другое слагаемое электростатической энергии, $-1.2 e^2 N/R$, описывает (в модели однородно распределённых в шаре элементарных электрических зарядов всех частиц) отрицательный вклад, происходящий от отсутсвия самодействия элементарных зарядов в квантовой механике. Потенциальная энергия $U(\mathbf{r})$ в уравнении Шредингера для заряженной элементарным зарядом частицы не включает потенциальную энергию взаимодействия рассматриваемой элементарной частицы с полем, которое она сама производит. Это значит что в квантовой механике действие частицы, заряженной элементарным зарядом, на саму себя исключается. Усредняя уравнение Шредингера квантово-механически получаем, что энергия частицы состоит из кинетической энергии и потенциальной энергии частицы во внешнем по отношению к ней поле. Так что видим, что энергия рассматриваемой частицы не содержит потенциальную энергию поля, создаваемого самой частицей, то есть в квантовой механике элементарный заряд $e$ в любом состоянии не обладает собственной потенциальной электростатической энергией (смотри, например [3]). Отсюда следует, что электростатическая энергия рассматриваемого нейтрального плазменного шара есть $-1{,}2 e^2 N/R$.

Для вырожденных электронов $g$ пропорционально $1/R^{1/2}$ [1,4]. Для классических электронов $g$ пропорционально $1/R^{3/2}$ [4]. Для $N$ частиц, как известно, отрицательный вклад энтропии в термодинамический потенциал есть $-3kTN\ln(R/R_0)$, где $R_0$ – некоторый параметр.

Для вырожденных электронов отрицательные слагаемые термодинамического потециала (одно из них пропорционально $1/R$, а другое пропорционально $1/R^{1/2}$, третье пропорционально $\ln(R/R_0)$) не могут скомпенсировать положительное слагаемое, кинетическую энергию, пропорциональную $1/R^2$, при уменьшении радиуса шара $R$. Так что, когда температура $T$ и плотность плазмы $N/V$ соответствуют вырождению делокализованных электронов, дальнейшее уменьшение размеров плазменного шара затруднительно.

Когда же температура достаточно высока и все частицы, как ядра так и делокализованные электроны, являются классическими, кинетичеcая энергия $2N$ частиц $K = 3kTN$; она не зависит от радиуса шара $R$. В этом случае дальнейшее уменьшение размера плазменного шара облегчается. Уменьшение размера плазменного шара приводит к увеличению плотности плазмы, что в свою очередь выводит плазму из области классической статистики, когда плотность плазмы возрастает. При этом дальнейшее уменьшение размера плазменного шара становится затруднительным. Для продолжения уменьшения размера плазменного шара (увеличения плотности плазмы) температуру и давление следует держать на должном уровне, когда имеет место классическая статистика. Этот процесс может привести к началу термоядерной реакции. Однако для такого развития событий требуются очень высокие температура и давление.

Следует заметить, что процесс уменьшения рамера плазменного шара в случае существенно вырожденных частиц возможен [5,6]. Здесь обсуждается возможность облегчения процесса уплотнения плазмы.

Литература